\documentclass[notitlepage,nofootinbib,
longbibliography,
amsmath,amssymb]{revtex4-1}

\usepackage{graphicx}  
\usepackage{dcolumn}   
\usepackage{bm}        
\usepackage{verbatim}   
\usepackage{bbold}
\usepackage{color}
\usepackage{xcolor}
\usepackage{amsthm}
\usepackage[normalem]{ulem}

\def\be{\begin{equation}}
\def\ee{\end{equation}}
\def\bea{\begin{eqnarray}}
\def\eea{\end{eqnarray}}

\def\>{\rangle}
\def\<{\langle}

\theoremstyle{definition}

\theoremstyle{remark}

\begin{document}

\title{Quantum oscillations in the activated conductivity in excitonic insulators: possible application to monolayer $\text{W}\text{Te}_2$.}

\author{Patrick A. Lee}
\affiliation{
Department of Physics, Massachusetts Institute of Technology, Cambridge, MA, USA
}

\date{October 19, 2020}
\begin{abstract}
A recent paper on the insulating state of monolayer $\text{W}\text{Te}_2$ reported the observation of large oscillations in the conductivity that are periodic in 1/B, resembling quantum oscillations in metals. This remarkable observation has inspired suggestions of exotic physics such as spin-charge separation. We show that a rather more conventional but still nontrivial explanation in terms of gap modulation may be possible in a model of excitonic insulator subject to a magnetic field.
\end{abstract}


\maketitle

\section{Introduction}
\noindent
A recent paper by Wang et al \cite{Sanfeng2020} reported the observations of quantum oscillations in the resistance in single layer $\text{W}\text{Te}_2$ when the system is gate tuned to an insulating state. This unexpected observation has led to speculations concerning possible exotic origins, such as the existence of fermionic excitons under the assumption that the holes have fractionalized into fermionic spinons and bosonic holons. In this paper we examine whether a more conventional explanation that does not appeal to spin-charge separation may be possible. We proceed in two steps. First we introduce a phenomenological model assuming that the energy gap of the insulator has oscillations that is periodic in 1/B as in conventional metals. We show that while the conductivity oscillation is exponentially suppressed by the gap, so is the background conductivity, so that the ratio between the two can be of order unity. Furthermore, at a temperature low compared with size of the gap oscillation, the conductivity exhibit large oscillations compared with the background, as if one is in the discrete Landau level regime in metals. This simple model can therefore explain the most remarkable observation made experimentally, i.e. the appearance of periodic sharp conductivity spikes at low temperature.

Next we show that a microscopic model of excitonic insulator indeed exhibit quantum oscillations of the energy gap. A recent paper by Jia et al \cite{Sanfeng_exciton2020} have shown evidence that the insulating state in monolayer $\text{W}\text{Te}_2$ is indeed an example of an excitonic insulator. We will follow their suggestion and provide a bit more microscopic details. Monolayer $\text{W}\text{Te}_2$ is known to be a semi-metal with overlapping conducting and valence bands. The valence band is located at the zone center while there are two electron pockets located along the $\Gamma$ to $Y$ direction, as shown in Fig 1. We will assume that both conduction and valence bands are described by isotropic quadratic dispersion about the band minimum and maximum. We will proceed in two steps. First we consider the case where there is a single electron band, so that at charge neutrality, the electron and hole have equal Fermi surface areas. We assume that Coulomb attraction between the electrons and holes leads to the formation of an excitonic insulator that gaps out the Fermi surfaces. Then we can make use of recent works \cite{knolle2015,fawang2016,knolle2017},that show that there are quantum oscillations in the physical properties with a period given the Fermi surface area before hybridization. Furthermore, these models indeed predict periodic in 1/B oscillation in the energy gap, even though its consequences for the resistivity has not been thoroghly explored. Next we consider the more realistic band-structure the consists of 2 electron pockets and describe the modifications that may introduce.

\begin{figure}[htb]
\begin{center}
\includegraphics[width=5in]{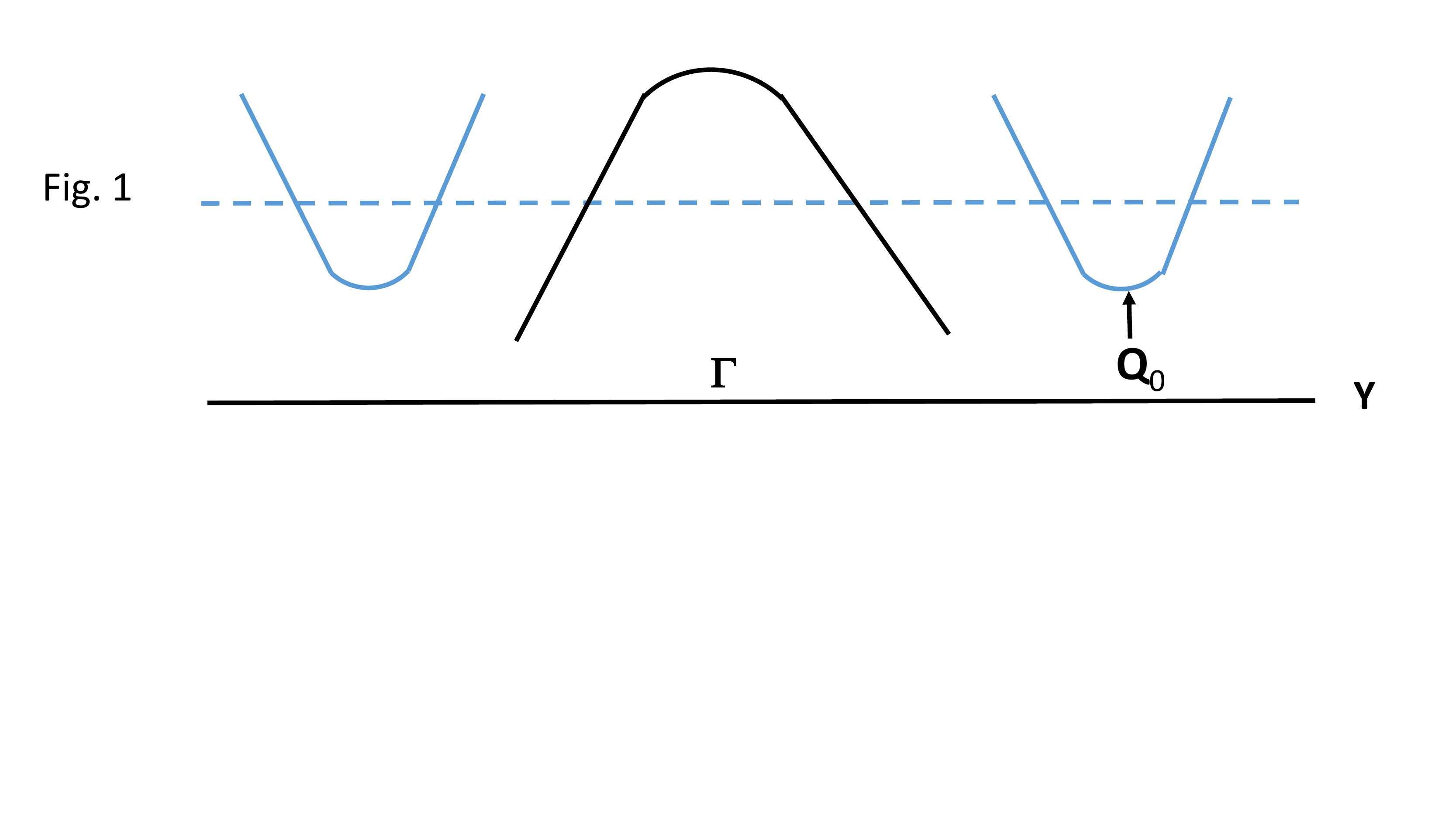}
\caption{Schematic band structure of $\text{W}\text{Te}_2$ showing the two conduction bands and a single valence band.}
\label{Fig: setup}
\end{center}
\end{figure}

\begin{figure}[htb]
\begin{center}
\includegraphics[width=6in]{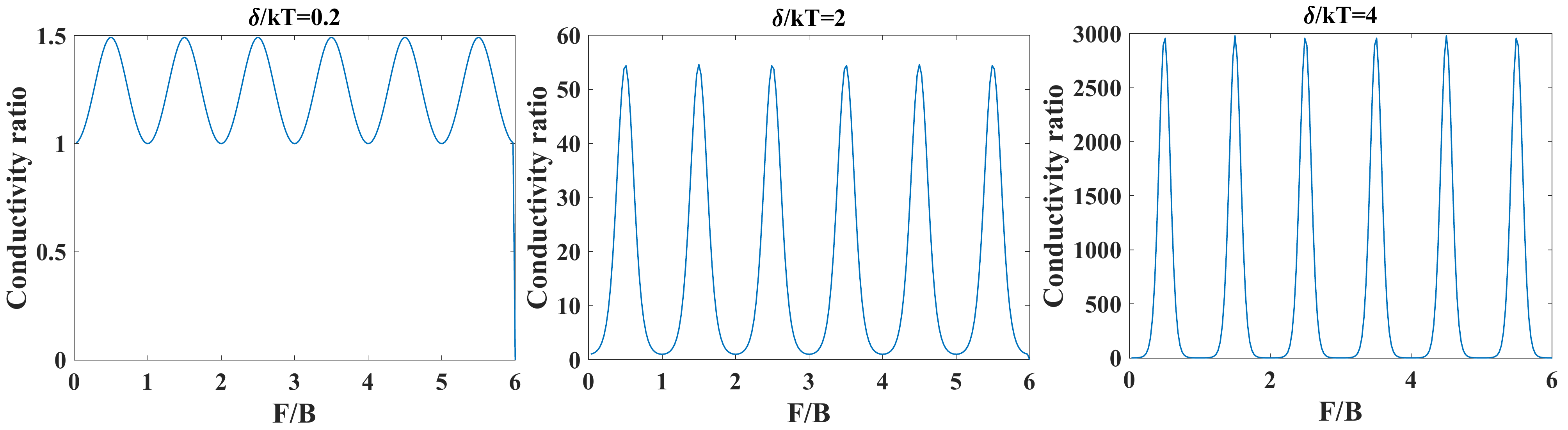}
\caption{Oscillation of the conductivity relative to the minimum value as a function of $ F/B$ for several values of $\delta/kT$}.
\label{Fig: dIdV1}
\end{center}
\end{figure}

\section{Phenomenological model of gap oscillations.}
\noindent
We assume a simple model of activated conductivity $\sigma$ with a gap $\Delta$ that is modulated periodically in $1/B$ with a period $F$ given by
\begin{equation}
    F=\hbar A/2\pi e
    \label{Eq: area}
\end{equation}
where $A$ is a Fermi surface area.
We further assume that the gap modulation is sinusoidal with an amplitude $\delta$.

\begin{equation}
\sigma = C e ^{-(\Delta_0  + \delta \cos(2 \pi F/B))/kT}.
\label{Eq: deltaD}
\end{equation}
For small  $\delta/kT$ we can expand to obtain 
\begin{equation}
\sigma / \sigma_0 = 1 - \delta /kT \cos(2 \pi F/B).
\label{Eq: ratio}
\end{equation}
where $\sigma_0 = C e^{-(\Delta_0 /kT)}$ is the average conductivity. A similar modulation obtains for the resistivity.  On the other hand, for large $\delta/kT$, it is better to consider the ratio to the minimum background conductivity $\sigma_0 ' = C e^{-((\Delta_0+\delta) /kT)}$ to obtain 
\begin{equation}
\sigma/\sigma_0' = e^{- \delta/kT (cos(2 \pi F/B)-1)}.
\label{Eq: ratio2}
\end{equation}
For $\delta/kT >> 1$ , this equation gives exponentially large spikes when plotted vs $1/B$ with width that is approximately $\sqrt{kT/\delta} / (2 \pi F) $.  Our simple model predicts sharp periodic spikes in both the conductivity and the resistivity in this regime. The addition of a small parallel constant conductivity will maintain the spikes in conductivity and eliminate the spikes in resistivity, in agreement with the experimental data.

In Fig 2 we plot the ratio $\sigma/\sigma_0'$ for several values of $\delta/kT$ and it is evident that the ratio evolves from small periodic modulations to large periodic spikes. Note that it is the ratio between the modulation amplitude $\delta$ and kT that controls the behavior, and not the average gap $\Delta_0$ itself.
Ref. \cite{Sanfeng2020} shows a cross-over from small oscillations to spike-like behavior when the temperature changes from 2K to 0.5K. The 0.5K data resembles our plot for $\delta/kT=4$, so we conclude that a gap modulation of about 2K is sufficient to give rise to these narrow conductivity peaks. Note that this is much smaller than the  gap scale estimated from the resistivity which is  about 12 meV or about 150K.

\section{The excitonic insulator model.}
\noindent

As mentioned in the introduction, monolayer $\text{W}\text{Te}_2$ is a semimetal according to band theory, and it is not clear where the energy gap comes from. A natural consideration is that the Coulomb attraction between the low density of electrons and holes lead to the formation of excitons, which then condense.\cite{Sanfeng_exciton2020} If the excitons are tightly bound, this is an example of Bose-Einstein condensation. On the other hand, for weaker coupling there is a state that is an analog of the BCS limit for superconductivity which is called the excitonic insulator. The analogy is clearest in the case of a single electron and hole band, both isotropic. We will first consider this simple case before discussion the more realistic band structure where there are two conduction bands and a single hole band. If the band extrema are separated by a momentum $\mathbf{Q_0}$, by spontaneously generating a density modulation at momentum $\mathbf{Q_0}$,  the electron and hole bands will hybridize at the Fermi level. Since there is perfect nesting at charge neutrality, a full gap will open at the Fermi level, giving rise to an insulator. The mathematics is very similar to BCS theory. The problem then maps to the problem of hybidized electron and hole bands considered in ref \cite{knolle2015,fawang2016,knolle2017}.  The idea is that the electron and holes retain memory of the Landau level in the presence of a perpendicular magnetic field. As a result, physical properties will exhibit quantum oscillations with a period given by the Fermi surface area of the bands before hybrization. So far, attention has been focused on thermodynamical properties such as magnetization, and the conclusion is that the cyclotron energy has to be comparable to the gap to get a sizeable effect.  On the other hand, properties such as conductivity that requires activation across the gap was briefly discussed in Ref. \cite{fawang2016}. Here we revisit this issue in greater detail.

Let us assume that the conduction and valence bands are given by the dispersion
$\epsilon_c(k)=k^2/2m_c + \epsilon_{c0}$ and similarly $\epsilon_v(k)=-k^2/2m_v + \epsilon_{v0}$ and $\epsilon_{v0}-\epsilon_{c0} > 0$. Here the electron band has been shifted by $\mathbf{Q_0}$ to intersect the valence band so that their Fermi surfaces coincide. We assume a hybridization matrix element $V$ which will open a gap at the Fermi level. For most of our discussion, we will ignore the spin degree of freedom. In the presence of a perpendicular magnetic field B, the momentum label is  replaced by Landau level indices $n$ so that the conduction band dispersion becomes $E_{c,n}=(\hbar eB/m_c c)(n+1/2)+ \epsilon_{c0}$ and similarly $E_{v,n}=-(\hbar eB/m_v c)(n+1/2)+ \epsilon_{v0}$ for the valence band. The energy eigenvalues are
\begin{equation}
E_{n,\pm}= 1/2(E_{c,n}+E_{v,n} \pm  \sqrt{(E_{c,n}-E_{v,n})^2+V^2})
\label{Eq: hybridize}
\end{equation}

As pointed out in ref \cite{ knolle2015,fawang2016},  the Landau levels with index $n$ that are closest to the Fermi level form the lowest energy state at the gap. The spectrum shows periodicity in 1/B with a period given by Eq. \ref{Eq: area} with the area $A$ given by the Fermi surface before the application of the magnetic field. As seen from Fig. 3  the gap edge shows clear modulation that is periodic in $1/B$. The thermal excition to the gap edge will dominate the conductivity, therefore justifying the phenomenological model introduced in section II. We note that ref \cite{fawang2016} assumed a bybridization which is linear in momentum and the Landau level $n$ of the conduction band is coupled to level $n+1$ or $n-1$ of the valence band. This leads to a complication that the gap is closed for small $1/B$. Since we assume a constant hybridization $V$ , the gap is always present.

\begin{figure}[htb]
\begin{center}
\includegraphics[width=6in]{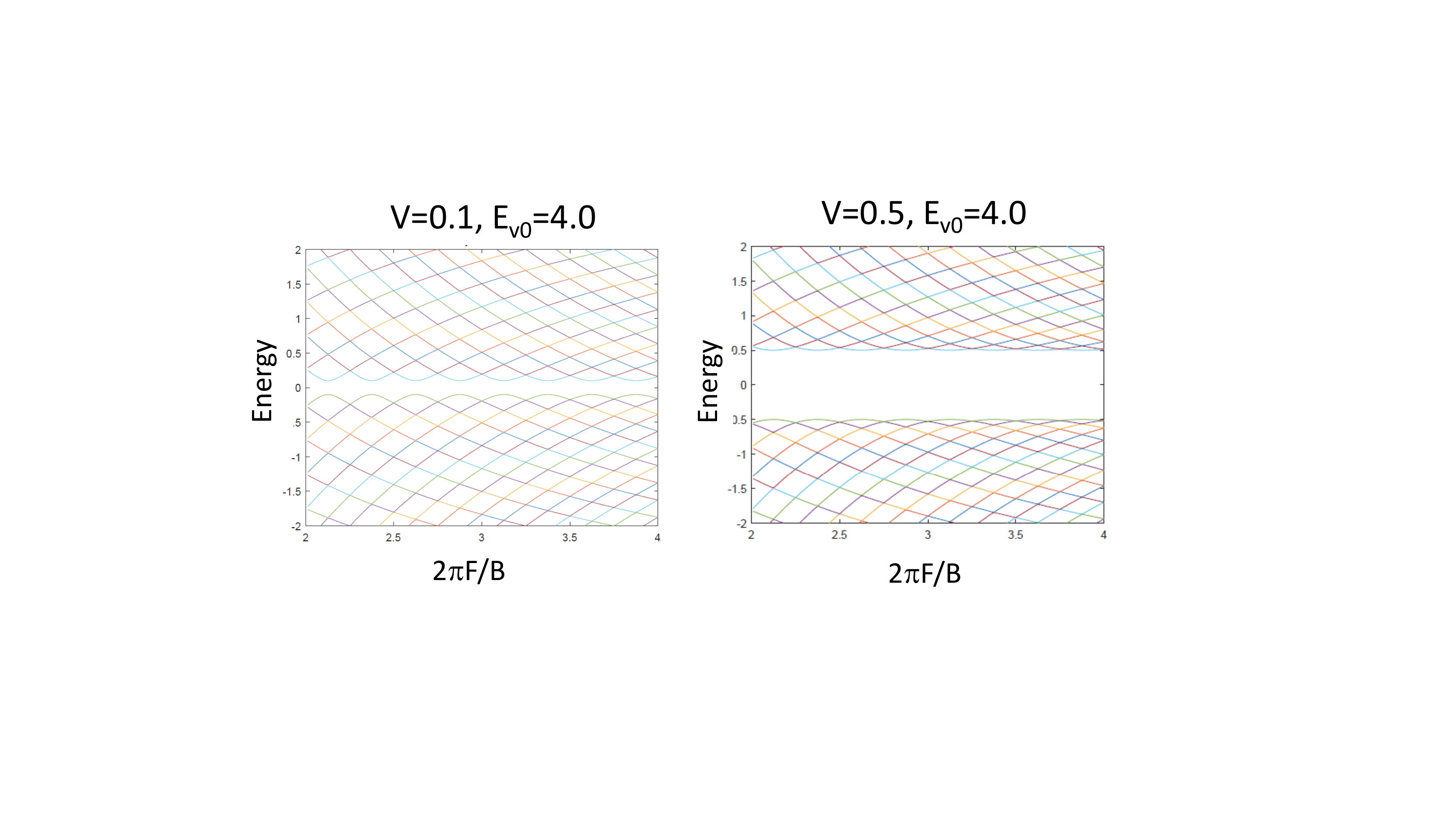}
\caption{Energy spectrum of the excitonic insulator as a function of $1/B$ for two values of hybridization $V$. As $B$ increases and $1/B$ decreases, Landau levels move up in energy for the conduction band and down in energy for the valence band. These are shown as the lines with negative and positive slopes respectively. At the Fermi level the Landau levels with the same index $n$ meet and hybridize, forming the gap. The gap is modulated in a periodic way and the relative size of the modulation decreases with decreasing $\hbar \omega_c / V $, where $ \omega_c = eB/m^*c$ and $1/m^* = (m_c+m_v)/m_c m_v$. This is a plot of Eq \ref{Eq: hybridize} where   $\hbar e /m^* $ has been set to unity. We have set $m_c=m_v$ and $E_{v0}=-E_{c0} =4$, so the period in $1/B$ is $1/E_{v0}=1/4$.}
\label{Fig: dIdV2}
\end{center}
\end{figure}

We can estimate the size of the band gap modulation by examining Eq. \ref{Eq: hybridize}. For $V \gg \hbar \omega_c$ where $\omega_c = eB/m^* c$ where $1/m^*=(m_c+m_v)/m_c m_v$, we can expand the energy as a function of $n$ about the gap minima and by setting the change in $n$ to 1/2, we estimate  the gap modulation to be approximately $(\hbar \omega_c)^2/8V$. This roughly corresponds to the parameter $\delta$ introduced in section II.

We note that ref \cite{fawang2016} also estimated the periodic component of thermally activated quantities and concluded that its effect is small. Their statement refers to the absolute magnitude of the oscillations. Our point is simply that the relative magnitude can be large.

  \label{Eq: DT1}.

  \label{Eq: DT2}  


Finally we comment on the applicability of these ideas to the case of $\text{W}\text{Te}_2$ where there are two conduction bands and a single valence band. If we simply introduce a potential at momentum $\pm \bf{Q_0}$ and translate the electron bands to overlap the valence band at the zone center, the bands cross above the chemical potential and a gap in general will not form at the Fermi level. This is true even for strong coupling, because an odd number of bands are being hybridized and a gapless band will generically remain. In order to get an insulating state, we propose the following two options.

First, we assume a self-consistent potential $V_2$ is generated which is at momentum $2\bf{Q_0}$ This will hybridize the two conduction bands. If $V_2$ is independent of momentum, this will simply split the conduction bands uniformly into symmetric and anti-symmetric  combinations. For sufficiently large $V_2$, only one band will cross the Fermi level, and the other one is empty. This problem now reduces to the simple model considered above. The period $F$ will be determined by the Fermi surface area of the  valence band in the absence of interactions and magnetic field. Ordinarily we expect that Coulomb repulsion between the conduction electrons will not favor a charge density wave, but may favor a spin density wave (SDW).   Here it is possible that strong Coulomb attraction with the holes will also help the electrons to stay close to each other in order to enjoy the attraction to the holes. A self-consistent theory involving both $V$ at momentum $\bf{Q_0}$ and $V_2$ at $2\bf{Q_0}$ will be required. On the other hand if SDW between the conduction electrons is favored,(since spin-orbit coupling is strong in $\text{W}\text{Te}_2$, the notion of spin should be replaced by pseudo-spin of time reversed partners)  the hybridization of the split band with the valence band may proceed as before, the only complication being that the hybridization matrix element is modified. 

\begin{figure}[htb]
\begin{center}
\includegraphics[width=6in]{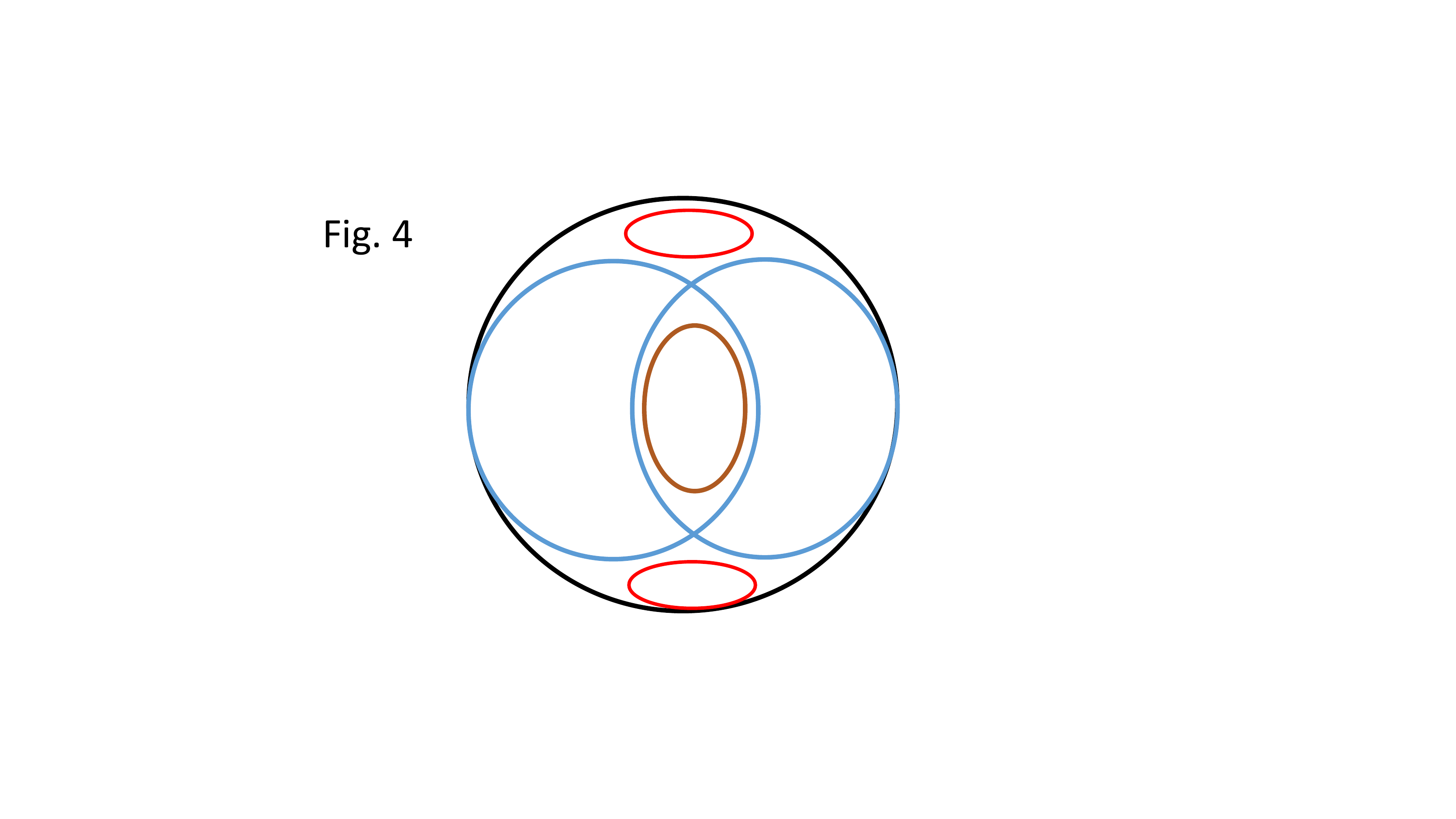}
\caption{Sketch of the conduction bands (blue)  Fermi surfaces after shifting the conduction bands to be tangential to the valence band Fermi surface (black). For modest hybridization, an electron pocket (brown) and two hole pockets (red) are left. With stronger hybridization, these will be squeezed out to form an insulator.}
\label{Fig: 2bands}
\end{center}
\end{figure} 

A second possibility is that a density wave at momentum $\pm \bf{Q}$ with a magnitude slightly larger than that of $\bf{Q_0}$ is generated, in such a way that the Fermi surfaces of the shifted conduction bands (blue) and the valence band are tangential at the Fermi level in order to optimize the nesting. This is shown in Fig 4. For weak coupling a hole pocket (red) and an electron pocket brown) will be left over, and the system remains metallic. In order to gap out the system we require either very strong coupling, or the introduction of a potential $V_2'$ at $2\bf{Q_0}$ which is effective in gapping out the crossing between the conductions bands (blue lines)  shown in fig 4. For a large enough $V_2'$, the electron and hole pockets will be squeezed out, resulting in an insulator. In the presence of a magnetic field, we can first consider the conduction bands subject to $V_2'$ large enough such that the overlap area is squeezed out and only a single Fermi surface remains. This Fermi surface will no longer be circular, but it has the same area as $A_v$ , i.e. that of the valence band in the absence of interaction. According to Onsager, discrete Landau levels will also form and the quantization condition will continue to be determined by the periodicity $F$ given by Eq. \ref{Eq: area}
with the area given by $A_v$, even though the wavefunctions will be much more complicated compared with the Landau wavefunctions. These discrete level will hybridize with the valence band Landau levels as before. The difference is that it is no longer true that only Landau levels with the same index $n$ will hybridize, but instead a set of matrix elements $V_{n,n'}$ will need to be calculated. We expect the structure of the eigenvalues as a function of $1/B$ will look similar to fig 2. It will show periodicity, but the size of the modulation cannot be estimated easily as before. This problem requires a numerical computation of the matrix elements and diagonalizing a large matrix, a task which is beyond the ability of this author.

In order to confirm the excitonic insulator picture, a direct measurement of the charge density modulation at a finite momentum by STM, for instance, will be highly desirable. Comparison of this momentum with the $\bf{Q_0}$ obtained from band structure will distinguish between the two scenarios.

\section{Conclusion.}
\noindent

 We have shown that large conductivity variations that mimic quantum oscillations can occur in an excitonic insulator where Coulomb attraction between partially occupied overlapping conduction and valence bands creates an insulator out of a semi-metal. When there is a single conduction and a single valence band, this conclusion follows very naturally from previous work with some slight modifications , but we argue that the picture may survive in the case where there are two conduction band and and a single valence band, as in $\text{W}\text{Te}_2$, The frequency of the oscillation is given by the Fermi surface area of the valence band in the absence of interaction. In ref. \cite{Sanfeng2020}, the frequency was reported to be 48.6T for sample 1 and 23T for sample 2. This translate to a Fermi surface area $A$ which we can write as $\pi k_F^2$ to extract a Fermi wave-vector $k_F \approx 4.8 \times 10^{-2} \AA ^{-1} $ for sample 1. This about $5 \% $ of the $\Gamma$ to $Y$ distance in the Brillouin zone and is consistent with the valence band Fermi wave-vector shown in the LDA band calculation. \cite{qian2014} Furthermore, in a third sample, the main peak in the Fourier spectrum  was found at around 25T. In this sample quantum oscillations are also observed in the electron and hole doped metals. The authors of ref.  \cite{Sanfeng2020} made the interesting point that the Fermi surface areas of the electron and hole metals extrapolate towards the 25T seen in the insulator.(See their extended data fig 3g.)  This is consistent with our picture, because the oscillation area is inherited from the electron and hole Fermi surfaces whose areas coincide at the charge neutrality point to form the excitonic insulator.

{\em Acknowledgment:}
I thank Sanfeng Wu for sharing his results with me prior to publication and to him and to N. P. Ong for helpful discussions.  This work has been supported by DOE office of Basic Sciences grant number DE-FG02-03ER46076.

\bibliographystyle{apsrev4-1}
\bibliography{test}
\end{document}